\documentclass{article}
\usepackage{spconf,amsmath,graphicx,hyperref}

\title{SAR-LM: SYMBOLIC AUDIO REASONING WITH LARGE LANGUAGE MODELS}
\name{Termeh Taheri, Yinghao Ma, Emmanouil Benetos}

\address{School of Electronic Engineering and Computer Science,\\
Queen Mary University of London, London, United Kingdom}

\begin{document}
\ninept
\maketitle
\begin{abstract}
Large language models (LLMs) have advanced in text and vision, but their reasoning on audio remains limited. Most existing methods rely on dense audio embeddings, which are difficult to interpret and often fail on structured reasoning tasks. Caption-based approaches, introduced in recent benchmarks such as MMAU, improve performance by translating audio into text, yet still depend on dense embeddings as input, offering little insight when models fail.

We present SAR-LM, a symbolic audio reasoning pipeline that builds on this caption-based paradigm by converting audio into structured, human-readable features across speech, sound events, and music. These symbolic inputs support both reasoning and transparent error analysis, enabling us to trace failures to specific features. Across three benchmarks, MMAU, MMAR, and OmniBench, SAR-LM achieves competitive results, while prioritizing interpretability as its primary contribution.

\textbf{Code available at:} \href{https://github.com/termehtaheri/SAR-LM}{github.com/termehtaheri/SAR-LM}
\end{abstract}
\section{Introduction}
\label{sec:intro}

Reasoning about sound is essential for building AI systems that can understand the world. While humans easily infer who is speaking, what their intention is, or what caused a noise, large language models (LLMs) still lag behind in this ability. Most prior work has evaluated LLM reasoning on text or vision \cite{xu2023largereasoning,yu2023sevila}, with audio reasoning only recently gaining traction. Current approaches typically rely on dense audio embeddings \cite{sakshi2024mmau}, which achieve moderate accuracy but are difficult to interpret and provide little insight into why models fail.

To improve interpretability, caption-based pipelines were proposed in the MMAU benchmark \cite{sakshi2024mmau}, where audio is first described in natural language before being reasoned over by an LLM. This improves performance but still relies on dense embeddings for caption generation, limiting transparency and error analysis.

In this work, we explore symbolic audio reasoning, where raw audio is translated into structured, human-readable features. We extract transcripts, speech emotion, sound events, musical notes, chords, and tags using pretrained and signal processing models. These features can be used directly or summarized into captions, and reasoning is then performed by an LLM. Because each symbolic layer is explicit, failure cases can be traced to specific components, making the pipeline more explainable than end-to-end captioning. We evaluate SAR-LM on three benchmarks: MMAU \cite{sakshi2024mmau}, MMAR \cite{ma2025mmar}, and OmniBench \cite{omnibench2024}, with further details provided in Sections~\ref{sec:related} and~\ref{sec:setup}.
 Our results show that symbolic reasoning enables detailed error analysis while achieving competitive accuracy, particularly in speech and environmental sound reasoning, with interpretability as the main contribution.
To the best of our knowledge, this is the first work to introduce symbolic audio reasoning with large language models. Our main contributions are:
\begin{itemize}
    \item We propose SAR-LM, a modular pipeline that converts audio into symbolic, interpretable features for reasoning.
    \item We evaluate SAR-LM on three benchmarks (MMAU, MMAR, OmniBench), showing competitive performance while improving interpretability.
    \item We demonstrate detailed error analysis enabled by symbolic features, exposing model failures that are hidden in embedding-based systems.
\end{itemize}

\section{Related Work}
\label{sec:related}

\subsection{Audio reasoning and the limits of current systems}
\label{sec:reasoning}

Audio reasoning moves beyond classification or captioning to answering questions that require inference from sound. Applications such as assistive agents or surveillance demand this capability, yet current systems remain limited.

Benchmarks such as MMAU \cite{sakshi2024mmau}, MMAR \cite{ma2025mmar}, and OmniBench \cite{omnibench2024} define the task across speech, music, and environmental audio, often requiring temporal reasoning. Early methods include Audio-CoT \cite{ma2025audiocot}, which applied chain-of-thought prompting \cite{wei2022chainofthought}, and Audio-Reasoner \cite{audio_reasoner2025}, which refined prompt structures with additional objectives.

Despite progress, most pipelines still depend on dense embeddings such as CLAP \cite{elizalde2022clap} or BEATs \cite{chen2023beats}, which are opaque to LLMs and weak at capturing temporal order. This leads to failures in tasks like event sequencing \cite{fayek2019daqa,sridhar2024temporal} and can trigger hallucinations when chain-of-thought is used on underspecified inputs \cite{xie2025audioreasoner}.

Follow-up work explores alternatives: Acoustic Prompt Tuning \cite{acoustic_prompt_tuning2025} with learnable prompts, Omni-R1 \cite{omni_r1_2025} showing metadata-only inference, Joint Audio-Speech Co-Reasoning \cite{joint_co_reasoning_2025}, and Stepwise Audio Reasoning \cite{stepwise_reasoning_2025}. These attempts highlight the need for interpretable, time-aligned inputs. Our work addresses this by replacing dense vectors with symbolic features and structured prompts, enabling clearer and more transparent reasoning.

\subsection{Audio captioning}
\label{sec:captioning}

Dense audio embeddings are effective for captioning and reasoning but remain opaque and hard to align with language models. Discrete features such as transcripts, tags, or note sequences offer a more interpretable alternative.

Audio captioning has been widely studied as a cross-modal task. Datasets like Clotho \cite{clotho_2020} and AudioCaps \cite{audiocaps_2019} enable evaluation, and end-to-end captioning models \cite{mei2021audio,zhang2023actual,tang2024extending} achieve strong accuracy but rely heavily on dense embeddings. Modular and retrieval-based systems such as EnCLAP \cite{kim2024enclap}, DRCap \cite{li2025drcap}, and SLAM-AAC \cite{chen2025slam} increase flexibility, yet still depend on opaque feature spaces. Few works attempt to use such interpretable inputs, and even fewer explore structured prompting.

This gap motivates our approach: using symbolic features directly or summarizing them into captions to support interpretability in audio reasoning, in contrast to the predominantly embedding-driven prior art.

\vspace{-10pt}
\begin{figure*}[t]
    \centering
    \includegraphics[width=\textwidth]{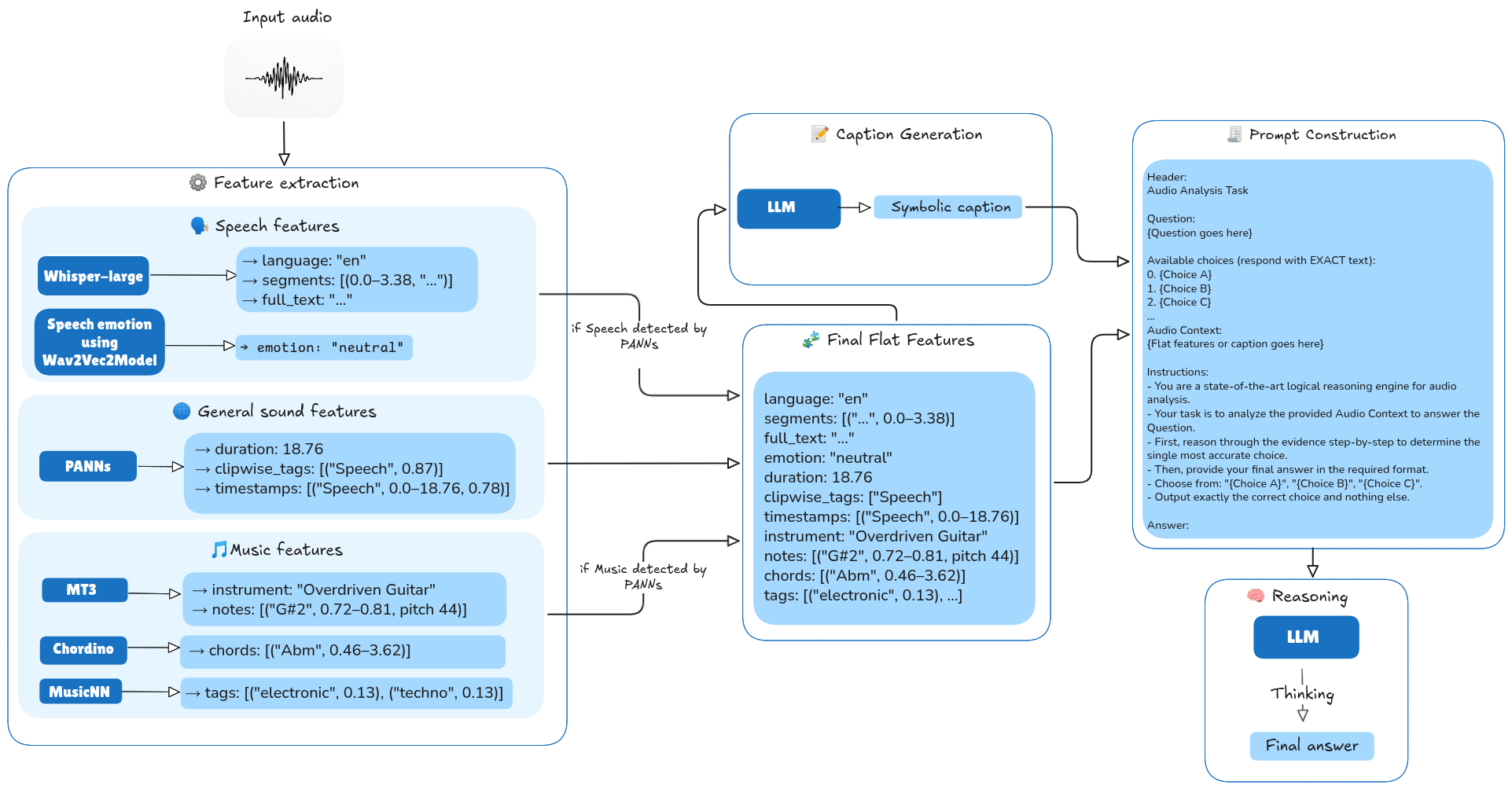}
    \caption{Overview of the SAR-LM pipeline. Raw audio is processed into symbolic features 
(speech transcription, sound event tags, music transcriptions), which are formatted into 
prompts and paired with benchmark questions. Symbolic features can also be summarized 
into captions before reasoning. Prompts are then reasoned over by an LLM to produce answers.}
    \label{fig:pipeline}
\end{figure*}
\vspace{-10pt}

\section{Methodology}
\subsection{Pipeline Overview}
\label{sec:pipeline}

Our goal is to help large language models (LLMs) understand and reason about audio. To achieve this, we design a modular pipeline that converts raw audio into structured, interpretable prompts for a language model. The pipeline consists of four main stages: symbolic feature extraction, prompt construction, LLM-based reasoning, and answer prediction, as shown in Fig.~\ref{fig:pipeline}.

Given an input audio clip $x$, we extract symbolic features using pretrained and signal processing models:
\vspace{-6pt}
\[
\mathcal{F}(x) = \{f_1, f_2, \dots, f_n\},
\]
where each $f_i$ is a discrete, time-aligned feature such as a transcript, tag, or chord sequence. These features are filtered and composed into a textual prompt $p = \mathcal{T}(\mathcal{F}(x), s)$, where $s$ denotes the selected prompt style. The prompt is paired with a question $q$ and passed to a large language model $\mathcal{M}$, which produces a predicted answer $\hat{y} = \mathcal{M}(p, q)$.

We support multiple prompt styles, including a flat format where all symbolic features are listed, and a caption-based format where the symbolic features are first summarized into a caption by the LLM before reasoning. The predicted answer is then evaluated against the ground-truth label from the benchmark.

The pipeline is fully modular and text-based: feature extractors, prompt generators, and reasoning models can be swapped independently without retraining. This design emphasizes interpretability, as symbolic features allow tracing successes and failures to specific components. Fig.~\ref{fig:pipeline} illustrates the full SAR-LM architecture.

\vspace{-6pt}
\subsection{Symbolic Feature Extraction}
\label{sec:features}

Instead of relying on dense embeddings, which are hard for LLMs to interpret, we convert each audio clip into symbolic, time-aligned features represented in text. These cover multiple semantic layers, sound events, speech, emotion, and music, with content-aware filtering to include only relevant features.

\textbf{Sound events.} We use PANNs \cite{panns_2020} to generate timestamped tags (e.g., footsteps, laughter, music), which both summarize the clip and guide which other modules are applied.

\textbf{Speech.} Whisper-large \cite{whisper_2023} provides full transcripts, while DAWN Transformer \cite{dawn_transformer_2023} predicts valence–arousal–dominance values that we discretize into symbolic emotion labels.

\textbf{Music.} MT3 \cite{mt3_2021} outputs symbolic note sequences with pitch, timing, and instrument information. Chordino \cite{mauch2009chordino,holloway_chord_extractor_2021} extracts chord progressions, and Musicnn\footnote{\url{https://github.com/jordipons/musicnn}} provides high-level tags describing genre or timbre.

All symbolic features are encoded as plain text and aligned to the audio timeline. They can be passed directly to the LLM or summarized into captions, enabling transparent reasoning and fine-grained error analysis.

\vspace{-10pt}
\subsection{Prompt Construction}
\label{sec:prompt}

Once symbolic features are extracted, we format them into a natural language prompt for the LLM. The goal is to present the audio content in a structured, interpretable way that supports reasoning.

We use PANNs as a reference to decide which features to include: if music is detected, we add notes, chords, and tags; if speech is detected, we add transcripts and emotion labels. This content-aware selection keeps prompts concise by including only relevant features.

All features are represented as plain text with a consistent structure. After them, we append the question, answer options, and an instruction block that specifies how the model should respond (e.g., select exactly one option).

Formally, the prompt $P$ is:
\vspace{-6pt}
\[
P = M \Vert F \Vert Q \Vert I
\]
where $\Vert$ denotes concatenation, $M$ is metadata (e.g., clip duration), $F$ is the selected feature set, $Q$ is the benchmark question with options, and $I$ is the instruction block.

\vspace{-6pt}
\subsection{Caption Generation}
\label{sec:caption}

To make symbolic features more human-readable and easier for LLMs to reason over, we also generate natural language captions. Symbolic features are reformatted into structured text and passed to a captioning model, which produces a fluent paragraph summarizing the audio scene. These captions provide an alternative abstraction level for prompting, complementing direct symbolic inputs and supporting interpretability in reasoning.
 
\vspace{-6pt}
\subsection{Reasoning with Language Models}
\label{sec:reasoning-llm}

We use large language models (LLMs) to answer multiple-choice audio questions based on text-only inputs. Prompts consist of either flat symbolic features or symbolic captions, followed by the benchmark question and options. 

Decoding constraints in the instruction block ensure stable outputs by limiting responses to a single selected option and reducing risks such as hallucination, overthinking, or token overflow.

\vspace{-6pt}
\section{Experiments and Results}
\vspace{-6pt}
\subsection{Setup}
\label{sec:setup}

We evaluated several large language models for both caption generation and reasoning. Qwen3-32B~\cite{qwen3technicalreport} often produced unstable outputs, sometimes overthinking even simple questions. Qwen3-30B-A3B-Instruct-2507~\cite{qwen3technicalreport} gave more reliable predictions, while Qwen2.5-Omni-7B~\cite{Qwen2.5-Omni} was tested as both captioner and reasoner but frequently hallucinated when generating captions from symbolic features. Based on these comparisons, we adopt Gemini~2.5~Pro~\cite{gemini25pro} as our final choice for caption generation and reasoning.

We evaluate on three benchmarks. MMAU~\cite{sakshi2024mmau} is a large-scale benchmark of 10k audio clips and 27 task types spanning speech, music, and environmental sounds, designed to test advanced perception and reasoning. We use the released mini-test set (1k samples) since its ground-truth labels are public. MMAR~\cite{ma2025mmar} provides 1k high-quality audio–QA pairs with multi-step chain-of-thought annotations, emphasizing deep reasoning across speech, music, and mixed audio. OmniBench~\cite{omnibench2024} extends evaluation to tri-modal settings, requiring reasoning over acoustic, visual, and textual inputs. Each dataset provides audio clips paired with multiple-choice questions, and we report reasoning accuracy overall and by category (speech, music, sound).

\vspace{-6pt}
\subsection{Overall Accuracy Across Datasets}

Table~\ref{tab:overall-results} compares three reasoning models on symbolic features across MMAU, MMAR, and OmniBench. 
Gemini~2.5~Pro achieves the highest accuracy on all datasets, with a clear advantage on MMAR (69.3\%). 
Qwen3-Instruct improves stability over Qwen2.5-Omni but still lags behind Gemini.

\vspace{-6pt}
\begin{table}[ht]
\centering
\caption{Overall reasoning accuracy (\%) across benchmarks using symbolic features.}
\label{tab:overall-results}
\resizebox{\columnwidth}{!}{%
\begin{tabular}{lccc}
\hline
\textbf{Reasoner} & \textbf{MMAU} & \textbf{MMAR} & \textbf{OmniBench} \\
\hline
Qwen2.5-Omni & 64.6 & 48.5 & 30.7 \\
Qwen3-Instruct & 67.2 & 54.9 & 35.8 \\
Gemini~2.5~Pro & \textbf{73.5} & \textbf{69.3} & \textbf{38.7} \\
\hline
\end{tabular}}
\end{table}
\vspace{-6pt}

\subsection{Dynamic Feature Selection with a GPT-Style Agent}

The symbolic pipeline involves many potential features, such as transcripts, sound events, chords, and music tags. Some of these are highly informative, while others may add noise depending on the clip. Manually testing the contribution of each feature is impractical, as the search space grows quickly. To address this, we introduced a GPT-style agent that selects relevant features per sample. Qwen2.5-Omni proved unstable in this role, often hallucinating selections, whereas Gemini~2.5~Pro provided consistent and meaningful choices. This dynamic selection helped reduce irrelevant inputs and improved downstream reasoning, as we demonstrate in the MMAU results.

\vspace{-6pt}
\subsection{Detailed Results on MMAU}

Table~\ref{tab:mmau-symbolic} reports task-wise and overall accuracy for different input formats. 
Captions are generated by Gemini~2.5~Pro, and the GPT-style agent for dynamic feature selection also uses Gemini~2.5~Pro. 
Reasoning is performed with Qwen3-30B-A3B-Instruct-2507 for controlled symbolic comparisons. 
We include both fixed symbolic inputs and dynamic agent-based selection. 
Flat symbolic features achieve strong performance in speech and sound reasoning, while symbolic captions perform slightly better on music. 
Agent-based selection provides consistent gains over non-agent variants, confirming that filtering out irrelevant features helps reduce noise. 
End-to-end captions are competitive for music tasks but remain less interpretable. 

\vspace{-8pt}
\begin{table}[ht]
\centering
\caption{Task-wise and overall accuracy (\%) on MMAU with Qwen3-Instruct as reasoner. Captions and agent-based selection use Gemini~2.5~Pro.}
\label{tab:mmau-symbolic}
\resizebox{\columnwidth}{!}{%
\begin{tabular}{lcccc}
\hline
\textbf{Input Type} & \textbf{Sound} & \textbf{Music} & \textbf{Speech} & \textbf{Overall} \\
\hline
Symbolic Features & 69.37 & 56.59 & \textbf{73.87} & 66.6 \\
Symbolic Features (agent) & \textbf{72.67} & 57.78 & \textbf{73.87} & \textbf{68.1} \\
Symbolic Captions & 69.67 & 58.38 & 71.77 & 66.6 \\
Symbolic Captions (agent) & 70.57 & 58.08 & 70.27 & 66.3 \\
End-to-End Captions & 68.17 & \textbf{62.28} & 69.97 & 66.8 \\
\hline
\end{tabular}}
\end{table}

\vspace{-10pt}
\begin{table}[ht]
\centering
\caption{Task-wise and overall accuracy (\%) on MMAU with Gemini~2.5~Pro as reasoner.}
\label{tab:mmau-gemini}
\resizebox{\columnwidth}{!}{%
\begin{tabular}{lcccc}
\hline
\textbf{Input Type} & \textbf{Sound} & \textbf{Music} & \textbf{Speech} & \textbf{Overall} \\
\hline
Audio-only & \textbf{80.18} & \textbf{72.46} & \textbf{83.18} & \textbf{78.6} \\
Symbolic Features & 73.27 & 64.97 & 82.28 & 73.5 \\
\hline
\end{tabular}}
\end{table}

We also evaluated Gemini~2.5~Pro directly as a reasoning model (Table~\ref{tab:mmau-gemini}). 
With raw audio as input, Gemini reaches the highest overall accuracy (78.6\%), showing the advantage of end-to-end access to the waveform. 
When reasoning over symbolic features, it still achieves 73.5\%, which is close in accuracy while providing far greater interpretability.

\vspace{-6pt}
\subsection{Comparison with Baseline Methods}

We compare our approach against reported baselines on the MMAU, MMAR, and OmniBench benchmarks. 
In all cases, we report our best-performing configuration: Gemini~2.5~Pro as reasoner with symbolic features as input. 
Results show consistent improvements over prior systems, particularly in speech reasoning and long-context MMAR tasks, while performance on OmniBench remains weaker, reflecting the added difficulty of tri-modal reasoning and highlighting an area for future improvement.

\vspace{-6pt}
\begin{table}[ht]
\centering
\caption{Comparison with baseline methods on MMAU (task-wise accuracy, \%).}
\label{tab:mmau-baseline}
\resizebox{\columnwidth}{!}{%
\begin{tabular}{lcccc}
\hline
\textbf{Method} & \textbf{Sound} & \textbf{Music} & \textbf{Speech} & \textbf{Overall} \\
\hline
MMAU (Best) & 57.35 & 49.70 & 64.86 & 57.30 \\
Audio-CoT & 62.16 & 55.99 & 56.16 & 58.10 \\
Audio-Reasoner & 60.06 & 64.30 & 60.70 & 61.71 \\
\hline
Ours (Gemini + Symbolic) & \textbf{73.27} & \textbf{64.97} & \textbf{82.28} & \textbf{73.5} \\
\hline
\end{tabular}}
\end{table}
\vspace{-6pt}

\begin{table}[ht]
\centering
\caption{Comparison with baseline methods on MMAR (task-wise accuracy, \%).}
\label{tab:mmar-baseline}
\resizebox{\columnwidth}{!}{%
\begin{tabular}{lcccc}
\hline
\textbf{Method} & \textbf{Sound} & \textbf{Music} & \textbf{Speech} & \textbf{Overall} \\
\hline
MMAR (Best) & \textbf{61.21} & 50.97 & 72.11 & 65.6 \\
Ours (Gemini + Symbolic) & 52.73 & \textbf{56.31} & \textbf{80.95} & \textbf{69.3} \\
\hline
\end{tabular}}
\end{table}
\vspace{-6pt}

\begin{table}[ht]
\centering
\caption{Comparison with baseline methods on OmniBench (task-wise accuracy, \%).}
\label{tab:omnibench-baseline}
\resizebox{\columnwidth}{!}{%
\begin{tabular}{lccc}
\hline
\textbf{Method} & \textbf{Sound} & \textbf{Music} & \textbf{Speech} \\
\hline
OmniBench (Best) & \textbf{60.00} & \textbf{52.83} & \textbf{55.25} \\
Ours (Gemini + Symbolic) & 29.43 & 48.11 & 40.60 \\
\hline
\end{tabular}}
\end{table}
\vspace{-6pt}

\subsection{Error Analysis}
\label{sec:error-analysis}

We illustrate a representative failure case in temporal reasoning.  
\textit{Question:} What was the order of the sounds in the sequence?  
\textit{Correct Answer:} light\_switch\_clicking $\rightarrow$ boiling\_water $\rightarrow$ doorbell\_ringing $\rightarrow$ clock\_ticking.  
Both flat symbolic features and symbolic captions predicted the wrong order, while the end-to-end captioner was correct.  

Closer inspection showed that PANNs failed to detect the early sounds (light switch, boiling water), leaving only later cues in the symbolic prompt.  
This missing context explains the error and illustrates a broader limitation: symbolic reasoning pipelines are only as reliable as their feature extraction.  
Unlike symbolic inputs, end-to-end captioners had full access to the waveform and recovered the correct sequence.  
Such interpretability is valuable, as it exposes where improvements are needed, in this case, more robust detection of short or low-energy sounds.

\section{Conclusion and Future Work}

We presented SAR-LM, a modular pipeline for symbolic audio reasoning with large language models.  
Our experiments across MMAU, MMAR, and OmniBench showed that symbolic inputs enable competitive reasoning performance while retaining interpretability that helps diagnose errors such as missed temporal events.  
Results highlight the potential of combining symbolic structure with strong reasoning models to outperform existing baselines, particularly in speech and long-context reasoning.

A limitation of this approach is that running multiple specialized models for feature extraction can be computationally expensive compared to end-to-end pipelines. 
In addition, performance is sensitive to the quality of symbolic features, as errors in speech recognition or music transcription can propagate to the reasoning stage. 
These trade-offs underline the need to balance efficiency with interpretability.

Future work will focus on strengthening the feature extraction stage, which remains the main bottleneck.  
One direction is to incorporate a universal sound recognition backend to reduce missed events from domain-specific models (e.g., PANNs).  
Another is to integrate stronger pretrained audio encoders such as MERT, fine-tuned per dataset, to capture richer and more generalisable features.  
Ultimately, unifying feature extraction across all audio types could yield both interpretability and improved accuracy, pushing symbolic reasoning systems beyond current end-to-end baselines.

\section{Acknowledgements}
Yinghao Ma is supported by the UKRI Centre for Doctoral Training in Artificial Intelligence and Music (EP/S022694/1).

\bibliographystyle{IEEEbib}
\bibliography{strings,refs}

\end{document}